\documentclass[twocolumn,showpacs,amsmath,amssymb,prl,superscriptaddress,floatfix]{revtex4}
\usepackage{graphicx}
\usepackage{color}

\begin{document}

\author{Andrea Micheli}
\affiliation{IQOQI and Institute for Theoretical Physics, Universities of Innsbruck, A-6020 Innsbruck, Austria}
\author{Zbigniew Idziaszek}
\affiliation{Institute of Theoretical Physics, University of Warsaw, Ho{\.z}a 69,
00-681 Warsaw, Poland}
\author{Guido Pupillo}
\affiliation{IQOQI and Institute for Theoretical Physics, Universities of Innsbruck, A-6020 Innsbruck, Austria}
\author{Mikhail A. Baranov}
\affiliation{IQOQI and Institute for Theoretical Physics, Universities of Innsbruck, A-6020 Innsbruck, Austria}
\author{Peter Zoller}
\affiliation{IQOQI and Institute for Theoretical Physics, Universities of Innsbruck, A-6020 Innsbruck, Austria}
\author{Paul S. Julienne}
\affiliation{Joint Quantum Institute, NIST and
the University of Maryland, Gaithersburg, Maryland 20899-8423, USA}

\title{Universal rates for reactive ultracold polar molecules in reduced dimensions}

\begin{abstract}
Analytic expressions describe universal elastic and reactive rates of quasi-two-dimensional and quasi-one-dimensional collisions of highly reactive ultracold molecules interacting by a van der Waals potential. Exact and approximate  calculations for the example species of KRb show that stability and evaporative cooling can be realized for spin-polarized fermions at moderate dipole and trapping strength, whereas bosons or unlike fermions require significantly higher dipole or trapping strengths.\end{abstract}


\maketitle

The successful production of relatively dense gases~\cite{Ni2008short} or lattices~\cite{Danzl2010short} of ultracold molecules in their ro-vibrational ground states opens up a number of new opportunities in physics and chemistry~\cite{Doyle2004short,Carr2009short}.   Reaction rates of ultracold molecules can be quite large, as measured and calculated for the fermionic species $^{40}$K$^{87}$Rb~\cite{Ospelkaus2010short,Idziaszek2010,Quemener2010}. This highly reactive species belongs to the class of molecules that have universal reaction rates that can be calculated analytically from a knowledge of the long-range van der Waals (vdW) potential alone, given unit probability of 
reaction at short range~\cite{Idziaszek2010}.  When an electric field is used to polarize a gas of this polar species, reaction rates become even larger and significantly limit the lifetime of the dipolar gas~\cite{Quemener2010,Ni2010short}.  A necessary condition for achieving stable dense samples of  such reactive ultracold molecules is to be able to understand and control such loss processes.  This is essential, for example, to achieve evaporative cooling in order to reach quantum degeneracy and thus utilize the molecules for new and exotic condensed matter phenomena that have been proposed~\cite{Micheli2006,Baranov2008,Lahaye2009short}.   Reference~\cite{Buchler2007short} predicted that molecular dipoles can be stabilized by tightly confining the molecules in a single plane of quasi-two-dimensional (quasi-2D) geometry. This is because the molecules can be arranged so as to experience repulsive dipolar forces at long range and thus never come together within chemical interaction distances in order to react.   First steps have been taken towards understanding these effects by recent quasi-2D calculations with adiabatic potentials ~\cite{Ticknor2010} or quantum dynamics for fermionic $^{40}$K$^{87}$Rb~\cite{Quemener2010b}.

Here we characterize quasi-2D and quasi-1D elastic and reactive collision rates of a broad class of bosonic and fermionic molecules that have universal rate constants.  Analytic expressions apply in the case of vdW interactions.  Quantum dynamical and approximate calculations for quasi-2D collisions of polar molecules show how collision rates scale with confinement length, dipole strength, and collision energy.  Our analysis connects the critically important experimental domain between the vdW and strong dipole limits, and provides a general framework for future research on molecular cooling, chemical reactivity, and condensed matter applications for species with universal collision rates.

Universal collision rates occur for chemical species that have near unit probability of reaction or inelastic relaxation when they are close enough together, on the order of typical chemical interaction distances, $a_c\lesssim 1$ nm, where strong chemical forces permit the reaction to occur~\cite{Idziaszek2010}.  Universal collision rates are completely determined by quantum threshold dynamics associated with the long range potential.  Consequently, there are two distinct classes of mixed alkali-metal diatomic species.  Those with energetically allowed reaction channels,  KRb, LiNa, LiK, LiRb, and LiCs~\cite{Zuchowski2010}, are expected to be universal.  By contrast, the species NaK, NaRb, NaCs, KCs, and RbCs have no reactive channels~\cite{Zuchowski2010}, and  are expected to be non-universal in their ground rotational, vibrational, and spin state.  However, even these species, as well as nonreactive homonuclear dimers, can have universal inelastic relaxation rates when vibrationally or rotationally excited~\cite{Julienne2009,Hudson2008short}.  Universal species do not have scattering resonances, since threshold bound states decay too fast to make one, whereas non-universal species can have numerous scattering resonances.  This paper treats the broad class of universal collisions.

\begin{figure}[tb]
	 \includegraphics[width=\columnwidth]{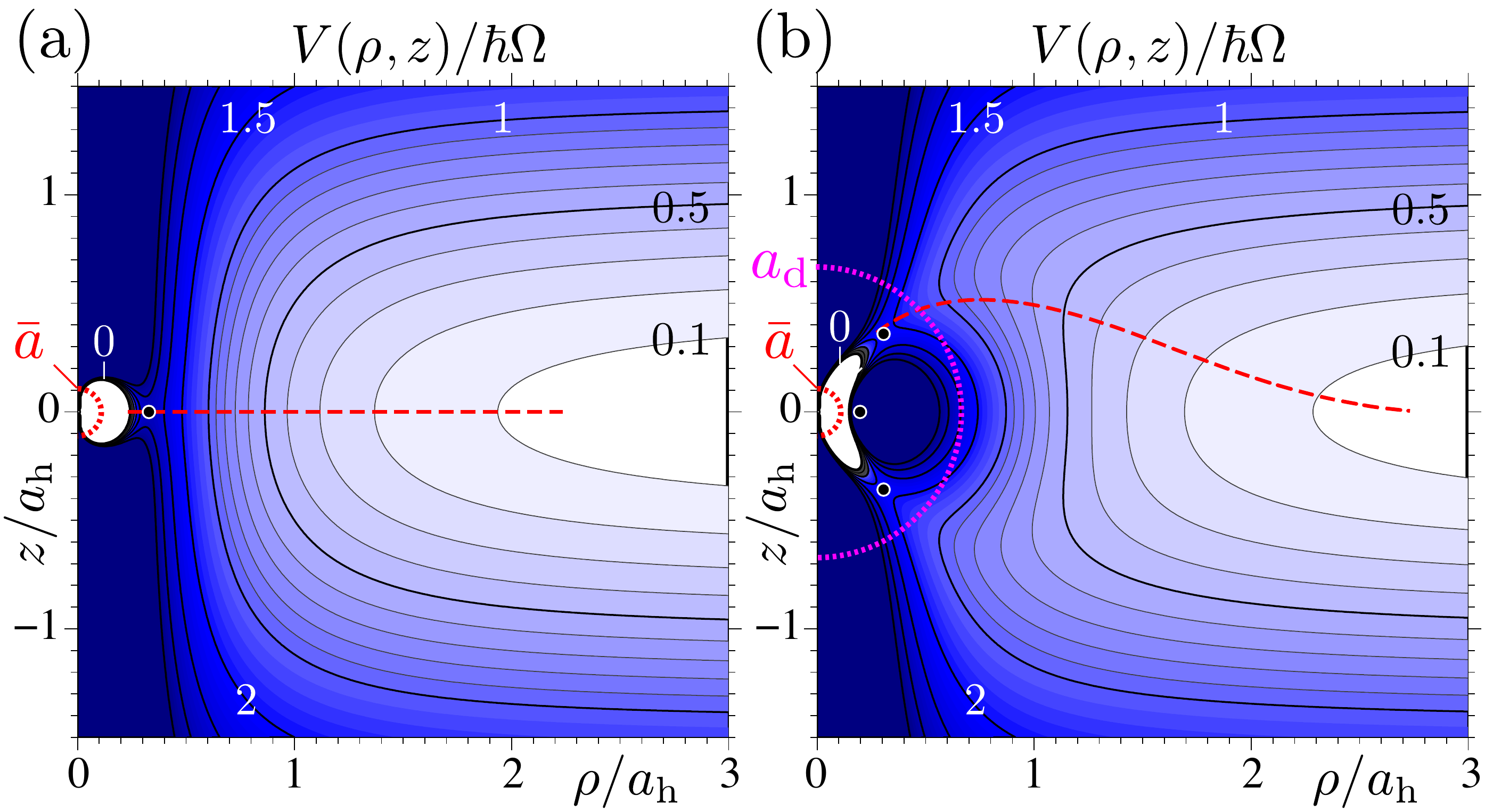}
	 \caption{\label{fig:1}Potential $V(\rho,z)/\hbar \Omega$ versus $\rho/a_{\rm h}$ and $z/a_{\rm h}$ for $|m|=1$ for $^{40}$K$^{87}$Rb in a $\Omega/2\pi=50\,\rm{kHz}$ trap, where $a_{\rm h}=56.4\,{\rm nm}$, $\bar{a}=6.25\,{\rm nm}$ with $C_6=16130\,{\rm a.u.}$~\cite{Idziaszek2010,Kotochigova2010}.  (a) vdW only case, $d=0$,  (b) dipolar case for $d=0.2\,{\rm D}$ (with $1\,{\rm D}=1\,{\rm Debye}=3.336\,10^{-30}\,{\rm C}{\rm m})$. Saddle points (white circles), (a) in-plane and (b) out-of-plane minimum action paths (dashed lines). Dotted half-circles indicate $\bar{a}$ and $a_{\rm d}$.}
\end{figure}

Figure~\ref{fig:1} shows the long-range potential $V(\rho,z)$ for two dipoles in quasi-2D geometry and illustrates essential features of reduced dimensional collisions, where
\begin{equation}
 V=\frac{\mu\Omega^2z^2}{2}+\frac{\hbar^2(m^2-1/4)}{2\mu\rho^2}-\frac{C_6}{r^6} +\frac{d^2}{r^3}\left(1-\frac{3z^2}{r^2}\right).
 \end{equation}
Here ${\bf r}=(\rho,\phi,z)$ represents the distance between the two molecules in cylindrical coordinates, and $r\equiv |{\bf r}|$.  The molecules are confined in the $z$-direction by a harmonic trap of frequency $\Omega$ and characteristic  length $a_{\rm h}=\sqrt{\hbar/\mu\Omega}$, where $\mu$ is the reduced mass of the pair.   The dipoles are assumed to be aligned along $z$, so the projection $m$ of their relative angular momentum is conserved.  The second term represents the $m$-dependent centrifugal potential in 2D. The third term is the isotropic vdW potential, assuming the molecules are in their rotational ground state, with a vdW length $\bar{a} = \left [2 \pi /\Gamma(1/4)^2\right ] \left(2 \mu C_6/\hbar^2\right)^{1/4}$~\cite{Gribakin1993,Julienne2009}. The last term is the anisotropic dipolar potential, with induced dipole moment $d$ and dipolar length $a_{\rm d}=\mu d^2/\hbar^2$.

Our model relies on the separation of length scales: $a_\kappa \gg a_{\rm h} \gg \bar a \gg a_c$, where $a_\kappa=2\pi/\kappa$ is the DeBroglie wave length for a collision with relative kinetic energy $E_\kappa= \hbar^2 \kappa^2/2\mu$. These inequalities are readily satisfied for experiments with  $^{40}$K$^{87}$Rb , for which $a_\kappa$, $a_{\rm h}$ and $\bar{a}$ can be on the order of hundreds of nm, tens of nm, and less than 10 nm, respectively. Under these conditions, collisions are essentially quasi-2D~\cite{Petrov2001,Li2008}, or if additional confinement at frequency $\Omega$ is provided along $x$, quasi-1D~\cite{Olshanii1998}.   Figure~\ref{fig:1}(a) illustrates the quasi-2D vdW case with $d=0$ for $|m|=1$, where  the minimal-action path followed by the two colliding particles lies in the plane $z=0$ of the contour diagram of $V(\rho,z)$. The competition between the centrifugal barrier and  the vdW attraction determines a single saddle point at a distance $r \gtrsim \bar a$, separating the long-distance 2D-scattering region from the short-range ``core,''  $r<\bar{a}$, where the molecules are accelerated towards one another by the attractive potential and experience 3D scattering. Two additional out-of-plane saddle points appear when $d$ increases so $a_{\rm d}/\bar{a} \gtrsim 2.71 (\bar a/a_{\rm h})^{3/2}$, as in Fig.~\ref{fig:1}(b), indicating a crossover to dipolar-dominated scattering, which is fully reached for $a_{\rm d} > a_{\rm h} \gg \bar a$. We show below that a strong enhancement of the ratio of elastic to reactive collisions in this regime will allow for an efficient cooling of the molecular gas.

Reference~\cite{Idziaszek2010} characterized elastic and reactive collision rates for 3D collisions by a complex scattering length $\tilde{a}_j(k)$, where $\hbar k$ is the 3D momentum.  For the special universal class of highly reactive molecules with unit short-range reaction probability, $\tilde{a}_0(k)=(1-i)\bar{a}$ for s-wave collisions of like bosons or unlike fermions, and $\tilde{a}_1(k)= (-1-i)(k\bar{a})^2\bar{a}_1$ for p-wave collisions of like fermions, with $\bar a_1\approx1.064\bar a$~\cite{Idziaszek2010}.  These explain the measured rates of 3D collisions of ultracold $^{40}$K$^{87}$Rb~\cite{Ospelkaus2010short,Idziaszek2010}.  Universal species have only incoming scattering current in the entrance channel in the vdW ``core'' of the collision, $r < \bar{a}$, and this provides a universal boundary condition for both the vdW and dipolar cases illustrated in Fig.~\ref{fig:1}.
While the general model can be extended to nonreactive species like RbCs, where one expects a rich resonance structure, we here focus on universal species like KRb.

Applying universal theory to quasi-1D and quasi-2D collisions in a vdW potential is straightforward for the $d=0$ case by combining the methods and notation of Refs.~\cite{Naidon2006,Idziaszek2010}.  We give here only the resultant formulas, valid for $\kappa \bar{a} \ll 1$.  Assume the molecule is prepared in its vibrational, rotational, and spin ground state and in the ground state of confined motion.  Only the first channel $j$ of a coupled channels expansion is needed near threshold for small $d$, and we set the index $j=0$ for like bosons or unlike fermions and $j=1$ for like fermions.  Elastic and reactive collisions in $N$ dimensions, $N=1,2,3$, are described by an $S$-matrix element $S_{jj}=\exp(i\theta_j)$ written in terms of a complex phase ${\theta_j}(\kappa)$ with
\begin{equation}\label{eq:taneta}
 \tan{\theta_j}(\kappa) = i\frac{1-S_{jj}(\kappa)}{1+S_{jj}(\kappa)} = -\tilde{a}_{j}(\kappa) \kappa^{N-2},
\end{equation}
where $\kappa=p,q,k$ represents the momentum in 1D, 2D, and 3D respectively~\cite{Naidon2006}.  The quantity $\tilde{a}_j(\kappa)$ on the right hand side defines the complex scattering phase through
\begin{equation}\label{eq:alpha}
\tilde{a}_j(\kappa)=\frac{L_j(\kappa)}{a_{\rm h}^{3-N}} \frac{(-1)^j(1+r_j)-i}{1+r_j+r_j^2/2}
\end{equation}
Table \ref{tab:1} gives the lengths $L_j(\kappa)$ and ratios $r_j(\kappa)$ for $N=1,2,3$, where $\xi_0=\bar{a}/a_{\rm h}$ and $\xi_1=\bar{a}_1\bar{a}^2/a_{\rm h}^3$. For the case in Fig.~\ref{fig:1}, $\xi_0=$ 0.111, $\xi_1=$ 0.00145, $q=(1/126)\,{\rm nm}^{-1}$, $q^2 a_{\rm h}^2=0.200$ for $E_\kappa=k_{\rm B}\,240\,\mathrm{nK}$, where $k_{\rm B}$ is Boltzmann's constant.  Assuming $\bar{a}/a_{\rm h} \ll 1$, we find $r_j\ll1$ and the right-hand factor in Eq.~\eqref{eq:alpha} can be approximated by $(-1)^j-i$, as in 3D~\cite{Idziaszek2010}.  The lengths $L_j$  change only weakly across dimensions $N$, so the scaling is mainly given by the factor $a_{\rm h}^{3-N}$.

\begin{table}
\caption{\label{tab:1}Lengths $L_j(\kappa)$ and ratios $r_j(\kappa)$ defining the universal complex scattering phase in $N$ dimensions;  $\zeta$ is the Riemann $\zeta$-function, $B = 0.905$ \cite{Petrov2001} and $W(0)=0.328$ \cite{IdziaszekPRL2006}.\\} 
\begin{tabular}{|r|c|c|c|}
\hline\hline
$N$&  \,\, 3D  \,\, &\,\, 2D \,\,   &  \,\, 1D \,\,  \\
\hline
$L_{0}$ & $\bar{a}$  & $\sqrt{\pi} \bar{a}$ & $2 \bar{a}$   \\
$L_{1}$ & $(k\bar{a})^2\bar{a}_1$  & $(3\sqrt{\pi}/2)(q\bar{a})^2 \bar{a}_1 $ & $6 (p\bar{a})^2 \bar{a}_1$   \\
$r_{0}$ &0 &  $(2/\sqrt{\pi})\xi_0\ln\left[2B/\pi q^2 a_{\rm h}^2\right]$ & $2\xi_0\zeta(1/2)$    \\
$r_{1}$ & 0 & $(2/\sqrt{\pi})\xi_1W(0)$ & $24\xi_1\zeta(-1/2)$    \\
\hline\hline
\end{tabular}\\
\end{table}

The elastic ${\cal K}_{j}^\mathrm{el}$ and reactive ${\cal K}_{j}^\mathrm{re}$ scattering rate constants follow from  the formulas in Ref.~\cite{Naidon2006}:
\begin{equation}\label{eq:RateK}
 {\cal K}_{j}^\mathrm{el}=\frac{\pi \hbar}{\mu}g_j \frac{ |1-S_{jj}|^2}{ \kappa^{N-2}},\quad
 {\cal K}_{j}^\mathrm{re}=\frac{\pi \hbar}{\mu } g_j \frac{1-|S_{jj}|^2}{\kappa^{N-2}} \,,
\end{equation}
where $g_0 = 1/\pi,2/\pi,2$ and $g_1 = 1/\pi,4/\pi,6$ for molecules colliding in like spin states in $N=1,2,3$ dimensions respectively, which take into account the $N$-fold degeneracy in $j=1$.  Unlike bosons or fermions have rate constants $({\cal K}_{0}+{\cal K}_{1})/2$.  The upper bound, or unitarity limit, on the rate constants follows immediately upon replacing the $S$-matrix expressions by their upper bounds, $|1-S_{jj}|^2\leq4$ and $ 1-|S_{jj}|^2\leq 1$, respectively. The elastic $\Gamma_{j}^\mathrm{el} = {\cal K}_{j}^\mathrm{el} n$ and reactive $\Gamma_{j}^\mathrm{re} = {\cal K}_{j}^\mathrm{re} n$ collision rates per particle are
\begin{subequations}\begin{align}
\Gamma^\mathrm{el}_{j}(\kappa) & = \frac{4\pi\hbar}{\mu} g_j L_{j}(\kappa) \frac{n}{a_{\rm h}^{3-N}} f_j(\kappa)
\eta_j(\kappa) \,,\label{Gamma_el}\\
\Gamma^\mathrm{re}_{j}(\kappa) & =  \frac{4\pi\hbar}{\mu} g_j L_{j}(\kappa) \frac{n}{a_{\rm h}^{3-N}} f_j(\kappa) \,, \label{Gamma_re}
\end{align}
\end{subequations}\noindent with $n$ the density of the collision partner in $N$ dimensions (units of cm$^{-N}$), $n/a_{\rm h}^{3-N}$ an equivalent 3D density (units of cm$^{-3}$) for $N=1$ or $2$ dimensions and $\eta_j(\kappa) = 2 L_{j}(\kappa) \kappa^{N-2}/a_{\rm h}^{3-N}$ gives the ratio of elastic to reactive collision rates.
The factor $f_j(\kappa) = \left[1+r_j(\kappa)+r_j(\kappa)^2/2 + \eta_j(\kappa) +  \eta_j(\kappa)^2/2\right]^{-1}$ approaches unity as $\kappa \to 0$, except for $j=0$ for $N=1,2$, where $f_0(p) \to\left[p a_{\rm h}/ \xi_0 \right ]^2/8$ and $f_0(q) \to \pi/2\ln^{2}\left[2 B/\pi q^2 a_{\rm h}^2\right]$, respectively.
However, for realistic traps and energies in the nK-regime the full expression for $f_0(q)$ is required in 2D, since the logarithmic term only becomes dominant at much lower energies.
 The expressions in Eqs.~\eqref{Gamma_el} and \eqref{Gamma_re} give explicitly the scaling with  $\kappa$ and $a_{\rm h}$, as well as all the known threshold law limits for collision rates in reduced dimensions~\cite{Sadeghpour2000short,Petrov2001,Olshanii1998,Li2008}.

We use both coupled-channels (CC) methods and analytic/semiclassical approximations to show the effect of the dipole moment on 2D collision rates.  The former uses a spherical harmonic basis set and a renormalized Numerov method to propagate the wave function $\Psi({\bf r})$ with universal incoming wave boundary conditions in the vdW core~\cite{Idziaszek2010}  out to distances $r \gg a_{\rm h}$. Then $\Psi({\bf r})$ is matched onto a cylindrical basis and propagated to larger $\rho$ to yield the 2D $S$-matrix.  The numerical results agree with the vdW limits from Eqs.~\eqref{eq:alpha}, \eqref{eq:RateK}, and Table~\ref{tab:1} and the dipolar results in Ref.~\cite{Quemener2010b}.

\begin{figure}[tb]
	\includegraphics[width=\columnwidth]{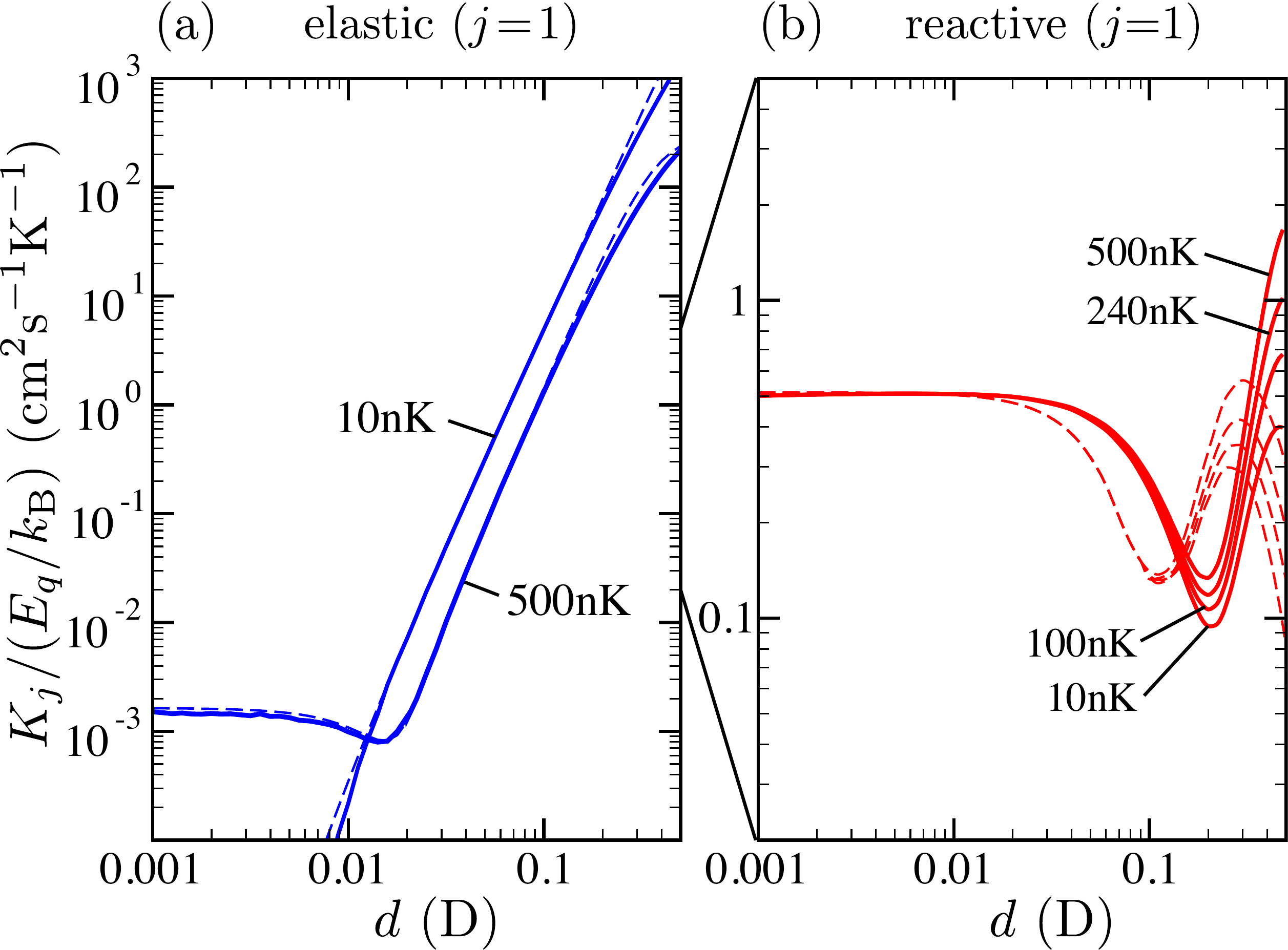}
	 \caption{\label{fig:2}Quasi-2D rate constants per energy, $K_1/(E_q/k_{\rm B})$, for elastic (a) and reactive (b) collisions of identical KRb fermions ($j=1$) versus dipole strength $d$ for an $\Omega/2\pi=50\,{\rm kHz}$ trap (cf. $a_{\rm d}=a_{\rm h}$ for $d=0.25\,{\rm D}$) and different energies $E_q/k_{\rm B}$; CC (solid lines) and UBA/instanton (dashed lines).}
\end{figure}

\begin{figure}[tb]
	 \includegraphics[width=\columnwidth]{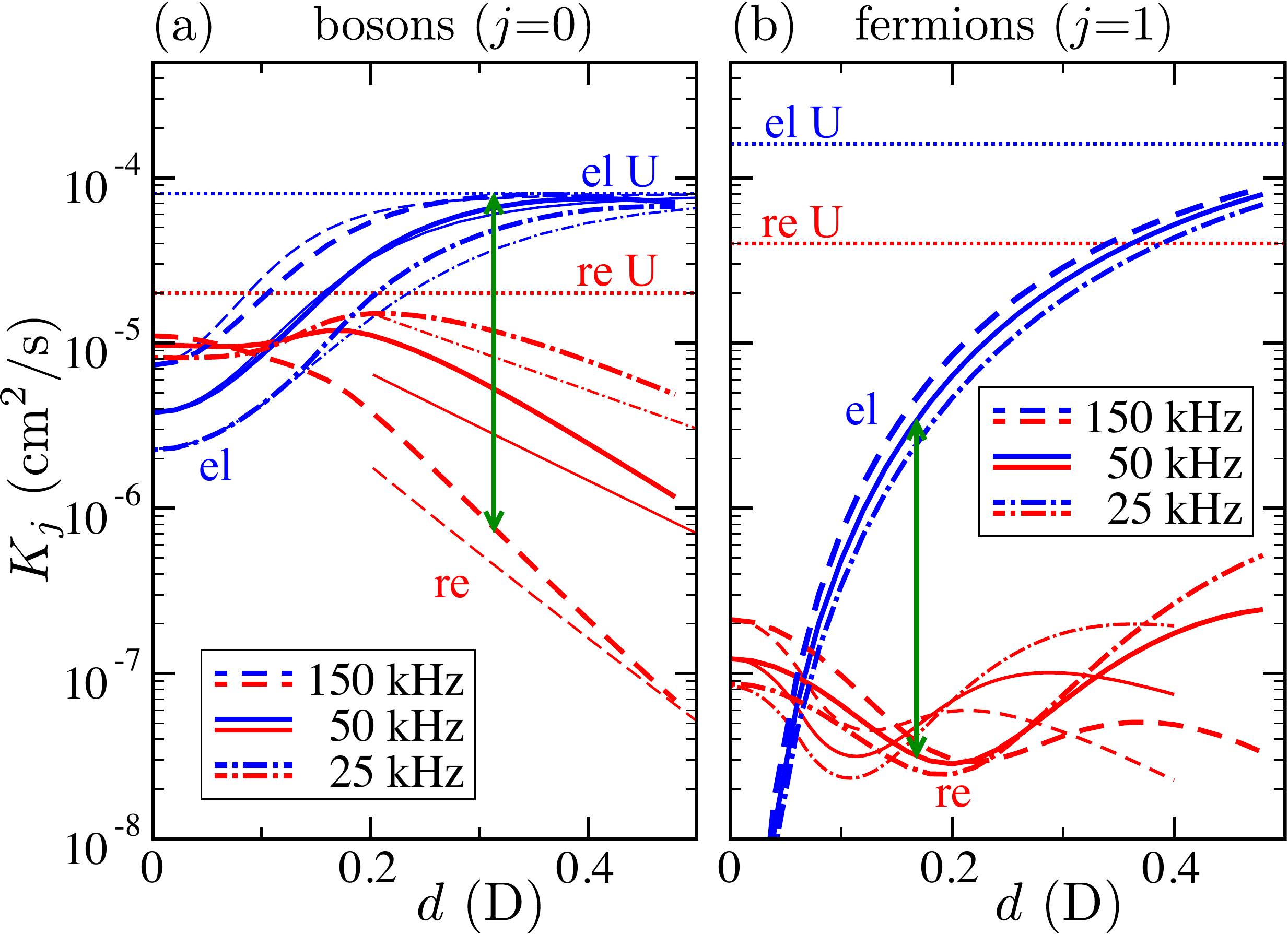}
	 \caption{\label{fig:3}Quasi-2D  elastic (el) and reaction (re) rate constants $K_j$ for  identical KRb (a) bosons ($j = 0$) and (b) fermions ($j=1$) at a collision energy of $E_q=k_{\rm B}\,240\,\mathrm{nK}=h\,5\,\mathrm{kHz}$ for three different trap frequencies $\Omega/2\pi=25\,{\rm kHz}, 50\,{\rm kHz}, 150\,{\rm kHz}$. CC (thick lines), UBA/instanton (thin lines), unitarity limits (dotted lines). Vertical arrows indicate where $\eta_j=100$.}
\end{figure}

Elastic collisions are well-described by a unitarized Born approximation (UBA), $S_{jj}^\mathrm{2D}=(1-iK_{jj}^\mathrm{2D})(1+iK_{jj}^\mathrm{2D})^{-1}$, where the $K$-matrix element in 2D includes the vdW term from Eq.~\eqref{eq:alpha} plus the dipolar term,
\begin{equation}
  K_{jj}^\mathrm{2D}(q) = -\tilde{a}_j^\mathrm{2D}(q)+2\sqrt{\pi}\frac{a_{\rm d}}{a_{\rm h}} \phi_j(qa_{\rm h})  \,,
  \label{UBA}
\end{equation}
with $\phi_0(x) = -0.65471+0.94146x -0.39010x^2 +{\cal O}(x^3)$ and $\phi_1(x) = -0.35555x +0.36042x^2-0.13417x^3  +{\cal O}(x^4)$.   Equation~\eqref{UBA} shows how the elastic rate constant scales with $q$, $a_{\rm d}$, and $a_{\rm h}$.  Figures~\ref{fig:2}(a) and \ref{fig:3}(a) show that  the UBA is an excellent approximation for dipolar bosons and fermions for a wide range of realistic $E_q$, $\Omega$ and $d$.

Reaction rates can be estimated using an instanton technique~\cite{Buchler2007short,Coleman1977} with $V(\rho,z)$ to get the transmission probability  $P_j^\mathrm{2D}$ for tunneling through the barrier separating the long-distance 2D-scattering region $r \gg \bar a$ from the short-range vdW core $r \lesssim \bar a$  (see Fig.~\ref{fig:1}):
\begin{equation}
P_j^\mathrm{2D}(q) = A_j e^{-S_{j}^{\rm cl}/\hbar}, \;\; S_{j}^{\rm cl} = 2\int_{{\bf r}_1}^{{\bf r}_2} ds \sqrt{2\mu[\tilde V_j({\bf r}^{\rm cl}_{j})-E_q]}. \nonumber
\label{instanton}
\end{equation}
Here, ${\bf r}_{j}^{\rm cl}$ is the path of minimal action given by the classical trajectory of a particle in the inverted potential with inner and outer turning points, ${\bf r}_{1}$ and ${\bf r}_{2}$, and $S_j^{\rm cl}$ the associated Euclidian action. We use $\tilde V_j=V_j+\hbar^2/8\mu\rho^2$ to take into account the semiclassical Langer-correction to the centrifugal term in the potential ($m^2-1/4\rightarrow m^2$), insuring correct threshold laws for $P_j^\mathrm{2D}(q)$.  We find $A_1 = 1.0297 (\bar a/a_{\rm h})$ by equating the analytic expression for $P_1^\mathrm{2D}(q)$ to the analytic $1-|S_{11}|^2$ in Eq.~\eqref{eq:RateK} in the vdW  limit $d \rightarrow 0$, assuming $A_j$ is independent of $d$ for $a_{\rm d}\lesssim a_{\rm h}$.  Since $S_{0}^{\rm cl}\to 0$  as $d\to 0$ due to the disappearance of a centrifugal barrier for $j=0$, we set $A_0=1$ to ensure unitarity is satisfied and only use $S_{0}^{\rm cl}$ to estimate ${\cal K}_{0}^\mathrm{el}$ at finite $d$ where a barrier exists and  $a_{\rm d}\lesssim a_{\rm h}$.

Figure~\ref{fig:2} shows that the variation of ${\cal K}_{1}^\mathrm{el}/(E_q/k_{\rm B})$ and ${\cal K}_{1}^\mathrm{re}/(E_q/k_{\rm B})$ with energy is relatively weak.      As in 3D, ${\cal K}_{1}^\mathrm{re}/(E_q/k_{\rm B})$ is independent of $E_q$ at low $d$,  but quite unlike in 3D~\cite{Quemener2010,Ni2010short}, it {\it decreases} relative to the vdW limit when $d$ increases.  The instanton method gives the qualitative explanation.  At small $d$ the barrier to the in-plane path increases with $d$, thus decreasing ${\cal K}_{1}^\mathrm{re}$.  As $d$ increases, the existence of out-of-plane saddle points gives alternative paths with a lower barrier, so ${\cal K}_{1}^\mathrm{re}$ starts to increase.  As $d$ increases more, the increasing out-of-plane barrier strength eventually  will cause $P_j^\mathrm{2D}(q)$ to again decrease with increasing dipole, evident in the CC and instanton 150 kHz  strong trap case in Fig.~\ref{fig:3}(b).  Comparing Figure \ref{fig:3}(a) and \ref{fig:3}(b) shows the instanton approximation gives the  qualitative trends for the boson case even better than for the fermion case at finite $d$.

Figure~\ref{fig:3} illustrates the stability and cooling properties expected for universal polar bosons and fermions.  In both cases ${\cal K}_{j}^\mathrm{el}$  changes  from the vdW limit by rapidly increasing with $d$ until it approaches the 2D unitarity limit at large $d$.   Stability requires that $\Gamma^\mathrm{re}_{j}(q)$ remain small enough that the lifetime $1/\Gamma^\mathrm{re}_{j}(q)$ is sufficiently long, order of $1\,{\rm s}$ or longer, as achieved for $^{40}$K$^{87}$Rb fermions in 3D~\cite{Ospelkaus2010short}.   Equation~\eqref{Gamma_re} predicts that the reaction rate per particle for identical fermions in the vdW limit is only $1/\sqrt{\pi}$ times lower in 2D than 3D, if the 2D system has the same equivalent 3D density $n^\mathrm{2D}/a_{\rm h}$.   The initial decrease in ${\cal K}_{1}^\mathrm{re}$ indicates not only increased fermionic stability in 2D, but also the possibility of evaporative cooling, which requires the ratio $\eta_j={\cal K}_{j}^\mathrm{el}/{\cal K}_{j}^\mathrm{re} \gg 1$.  Figure \ref{fig:3}(b) shows that $\eta_1$ reaches a magnitude near 100 for $d=0.18\,{\rm D}$, nearly  independent of the trap $\Omega$.  Figure \ref{fig:2} shows that the $\eta_1$ ratio increases with lower $E_q$, indicating that the evaporation improves as the 2D gas cools.

Figure~\ref{fig:3}(a) shows that there is less room to improve the bosonic elastic $\Gamma^\mathrm{el}_{0}(q)$ with increasing $d$, since it is only an order of magnitude below unitarity in the vdW limit.  The reactive $\Gamma^\mathrm{re}_{0}(q)$ at low dipole strength is much larger than for fermions, so universal polar bosons have shorter lifetimes and poorer stability than fermions at the same dipole and trap strength.  Furthermore, getting $\eta_0 \gtrsim 100$ requires either large $d$ or large $\Omega$.  Figure~\ref{fig:3} shows that $\eta_0=100$ in a 50 kHz trap near $d=0.5$ D or in a 150 kHz trap for $d=0.3$ D.  Thus, stability and evaporation for universal polar bosons may be achievable.

In conclusion, we have developed universal analytic expressions for quasi-2D or quasi-1D collisions of highly reactive ultracold bosonic or fermionic molecules in the absence of an electric field and have calculated quasi-2D collisions of universal polar molecules with a dipole moment.  While prospects for stability and evaporative cooling in 2D are much better for universal polar fermions than bosons,  either species would benefit from larger dipole moments or tighter confinement.  Non-reactive, non-universal species are expected to be very different from universal ones and to show numerous shape or Feshbach resonances as electric or magnetic fields are tuned.  

We acknowledge support from an AFOSR MURI, the EOARD, a Polish Government Grant for 2007-2010, the University of Maryland NSF-PFC, the Austrian FWF, and the EU NAME-QUAM.  We thank G. Qu\'em\'ener, J. L. Bohn, and Jun Ye for discussions.

\bibliography{mol2D}

\end{document}